\documentclass{article}
\usepackage{spconf,amsmath,graphicx}

\title{PLDA-based Diarization of Telephone Conversations}

\makeatletter
\def\name#1{\gdef\@name{#1\\}}
\makeatother
\name{{\em Ahmet Emin Bulut$^{1,2}$, Hakan Demir$^1$, Yusuf Ziya I{\c{s}}{\i}k$^{1,2}$, Hakan Erdogan$^2$}}

\address{$^1$TUBITAK BILGEM , Gebze, Turkey  \\
$^2$Faculty of Engineering and Natural Sciences, Sabanci University, Turkey\\
{\small \tt \{ahmet.bulut,hakan.demir,yusuf.ziya\}@tubitak.gov.tr, haerdogan@sabanciuniv.edu} }

\begin{document}
%
\maketitle
\begin{abstract}


This paper investigates the application of the probabilistic linear discriminant analysis (PLDA) to speaker diarization of telephone conversations. We introduce using a variational Bayes (VB) approach for inference under a PLDA model for modelling segmental i-vectors in speaker diarization. Deterministic annealing (DA) algorithm is imposed in order to avoid local optimal solutions in VB iterations. We compare our proposed system with a well-known system that applies k-means clustering on principal component analysis (PCA) coefficients of segmental i-vectors. We used summed channel telephone data from the National Institute of Standards and Technology (NIST) 2008 Speaker Recognition Evaluation (SRE) as the test set in order to evaluate the performance of the proposed system. We achieve about 20\% relative improvement in Diarization Error Rate (DER) compared to the baseline system.
\end{abstract}
\section{Introduction}
\label{sec:intro}

Nowadays, with the explosive growth of audio documents, there is an increasing interest towards applying speech technologies to automatic searching, indexing, and retrieval of audio information. Speaker diarization, which gives the ``who spoke when" information without any prior knowledge about speakers, is an important sub-task to address mentioned problems. To illustrate, for an automatic speech recognition system such information allows us to determine the occurrences of specific speaker for a given utterance, which in turn improves transcription performance by speaker adaptation. Moreover, successful diarization of conversations would also increase the performance of speaker verification systems. Speaker diarization of audio data has been studied for different domains, such as meeting, broadcast and telephone recordings \cite{sun2010meeting, barras2006broadcast, kenny2010diarFA}. 

Basically speaker diarization consists of three stages. In the first step, speech activity detection is employed in order to extract speech containing parts from a given utterance. As the second step, the extracted speech parts are further divided into segments according to the speaker changes in such a way that each segment contains the speech of a single speaker. This stage is called speaker segmentation in the literature. Finally, in the clustering stage, all the segments are passed over and the ones spoken by the same speaker are labeled identically. Speaker-based clustering can also be followed by cluster re-combination, which refines the speaker clusters for more purity. Among all the components of a speaker diarization system, performance of clustering stage is crucial for the success of the overall system. Many systems have been designed and tuned based on Bayesian Information Criterion (BIC). One such system \cite{mit2004diarization}, developed by MIT Lincoln Laboratory, serves as a baseline for a number of studies.

Upon the recent successes of factor analysis based methods, this study explores a new set of such approaches to speaker diarization. We adapt the methods from speaker recognition in order to make use of the concept of inter-speaker variability for the diarization of telephone conversations. Factor analysis based speaker diarization was first introduced in \cite{castaldo2008stream} using a stream-based approach. In the study of Kenny et al. \cite{kenny2010diarFA}, they modify Valente's \cite{valente2005varbayes} speaker diarization system based on the VB method and they incorporate the factor analysis priors defined by eigenvoices and eigenchannels \cite{kenny2008intervar}. Also, in a recent study \cite{prazak2011clusterPLDA}, PLDA is introduced to the problem of speaker diarization. They use factor analysis to extract low-dimensional representation of a sequence of acoustic feature vectors, namely i-vectors \cite{dehak2011frontFA} which are modelled by PLDA. As the metric for clustering, they use log-likelihood ratio of the probability of hypothesis that two clusters represented by corresponding i-vectors share the same identity and have distinct identities, rather than BIC-based clustering as used in \cite{mit2004diarization}. The authors in \cite{shum2011intraconv} proposed k-means clustering for i-vector based diarization approach which constitutes our baseline system. We also extract i-vectors for each segment in a similar way, however we represent i-vectors with a PLDA model and use a VB approach for inference under the model \cite{bishop2007}.
 
The rest of the paper is organized as follows. Section \ref{sec:PLDA_SD} provides the overview of our speaker diarization system. The experimental setup and results are then described in section \ref{sec:exp}. Section \ref{sec:conc} is devoted to conclusion and future work, and relation to prior works is explained in section \ref{sec:prior}.

\section{PLDA-based Speaker Diarization System}
\label{sec:PLDA_SD}

PLDA is originally used for the face recognition task \cite{prince2007PLDA}. Later, it is successfully applied to the speaker detection task as well \cite{brummer2010system, campell2010system}. In our study, PLDA is adapted to the speaker diarization problem by proposing a special generative story for segment i-vectors. This is the first study, to the best of our knowledge, PLDA is used for modelling the extracted segment i-vectors and inference under the model is realized by VB for speaker diarization.

Our speaker diarization system is composed of mainly three parts. Speaker change point detection, alignment of segments over speakers, and re-segmentation. The implementation details of the first and the last parts are similar with the earlier study in \cite{mit2004diarization}. For the second part, where we assign segments to speakers, we follow a VB approach with different initialization methods and a DA variant of VB \cite{katahira2008DAVB}.

\subsection{Two Covariance PLDA Model}

The i-vector features, contain information relevant to factors like channel, microphone, speaking style, language in addition to speaker identity. In speaker verification, PLDA model is used to extract speaker identity related factors from i-vectors. A variant of PLDA, known as two covariance PLDA \cite{brummer2010twocov}, assumes that the i-vectors are generated by addition of two terms; a speaker vector $y$ unique to a speaker and a residual vector $\epsilon$ unique to the utterance. The speaker vector $y$ is assumed to be sampled from a Gaussian distribution with mean $\mu$ and covariance $\Lambda^{-1}$, and the residual vector is assumed to be sampled from a Gaussian with zero mean and covariance $\mathcal{L}^{-1}$. 

\subsection{Modelling Assumptions}

We assume that we have a two covariance PLDA model trained on a separate training set at hand. We assume that we are given a conversation involving $S$ speakers and the speaker change points are specified. Let us denote the set of segment i-vectors by ${\bf\Phi}=\{\phi_1, ..., \phi_M\}$. For each segment $m = 1, ... , M$, we define an $S\times1$ indicator vector $i_m$ whose components are defined as $i_{ms}=1$ if speaker $s$ is talking in the segment $m$ and $i_{ms}=0$ otherwise. Let ${\bf I}=\{i_1,...,i_M\}$ be the set of all indicator vectors belonging to the given utterance. We also assign a prior probability to the event that a speaker $s$ is talking in a given segment; we denote and set by $\pi_s=\frac{1}{S}$. The generative story for our PLDA based diarization model is as follows:

\begin{itemize}
\item For each speaker $s$ sample $y_s$, from $\mathcal{N}(y;\mu,\Lambda^{-1})$.
\item For each segment:
    \begin{itemize}
    \item Sample $i_m$ from the multinomial distribution $Mult(\Pi)$ where $\Pi=(\pi_1,...,\pi_S)$. Let $k$ be the index for which $i_{mk}=1$, with all the other entries of $i_m$ being 0.
    \item Sample $\epsilon_m$ from $\mathcal{N}(\epsilon;\bar{0},\mathcal{L}^{-1})$.
    \item The observed segment i-vector is obtained as $\phi_{m}= y_k + \epsilon_m$.
    \end{itemize}
\end{itemize}
Let ${\bf Y}=\{y_1,...,y_S\}$ be the set of speaker vectors of the speakers talking in the given utterance. Using this model, we can summarize the diarization problem as of calculating the posterior probability of the speaker talking in a given segment. With these assumptions, obtaining the posterior probability, $P({\bf Y},{\bf I}|{\bf\Phi})$ produce intractable integrals. Therefore we resort to the  approximate inference methods, namely mean-field VB, in order to approximate $P({\bf Y}|{\bf\Phi})$ and $P({\bf I}|{\bf\Phi})$. 

\subsection{Variational Bayes for PLDA based i-vectors}
\label{vb_for_PLDA_ivec}

The basic assumption for mean-field variational methods is that the approximate posterior factorizes as:
\begin{eqnarray}
\label{Q_fact}
Q({\bf Y},{\bf I})=Q({\bf Y})Q({\bf I})
\end{eqnarray}
Approximate segment and speaker posteriors, $Q({\bf I})$ and $Q({\bf Y})$, are defined as:
\begin{eqnarray}
\label{Q(I)}
Q({\bf I}) = \prod_{m=1}^{M} \prod_{s=1}^{S}q_{ms}^{i_{ms}}
\end{eqnarray}
\begin{eqnarray}
Q({\bf Y}) = \prod_{s=1}^{S}\mathcal{N}(y_s\mid \mu_s,C_s^{-1})
\label{Q(Y)}
\end{eqnarray}
In equation (\ref{Q(I)}), we define $q_{ms}$ as the posterior probability of speaker $s$ talking in segment $m$ and in equation (\ref{Q(Y)}), it turns out that approximate speaker posterior distributions are Gaussian with mean $\mu_s$ and precision $C_s$.

Adapting the formulation in \cite{kenny2008BAdiar}, we formulate segment and speaker posteriors for the VB approach as follows:
\begin{enumerate}
\item Update rule for segment posteriors:
\begin{eqnarray}
q_{ms}=\frac{\tilde{q}_{ms}}{\sum_{s'=1}^{S}\tilde{q}_{ms'}}
\label{q_ms_norm}
\end{eqnarray}
where
\begin{eqnarray}
\begin{split}
\log \tilde{q}_{ms}= &\mu_s^T\mathcal{L}\phi_m-\frac{1}{2}\mathrm{tr}(\mathcal{L}(C_s^{-1}+\mu_s\mu_s^T)) \\
                     &+\mbox{const}
\end{split}
\label{q_ms}
\end{eqnarray}
where $\mbox{const}$ stands for speaker independent terms.

\item Update rule for speaker posteriors:
\begin{eqnarray}
C_s=\Lambda+\sum_{m=1}^{M}q_{ms}\mathcal{L}
\label{C_s}
\end{eqnarray}
\begin{eqnarray}
\mu_s=C_s^{-1}(\Lambda\mu+\sum_{m=1}^{M}q_{ms}\mathcal{L}\phi_m)
\label{mu_s}
\end{eqnarray}
\end{enumerate}
\newpage
The speaker and segment posteriors are updated alternately throughout the variational e-step. On convergence, diarization is performed by assigning each segment $m$ to the speaker given by $\underset{s}{\operatorname{argmax}} ( q_{ms})$ \cite{kenny2010diarFA}.

Initializing the VB algorithm by just assigning random values to the segment posteriors $q_{ms}$ is proved to be ineffective especially for the recordings that one speaker dominates the conversation \cite{kenny2010diarFA}. For that recordings, two speaker posteriors found by the VB algorithm only model the dominant speaker, and the diarization error rate may be very high corresponding to the average. In order to overcome this problem we try various initialization heuristics for a better start up for the VB iterations and also use a DA variant of the variational algorithm to avoid local optimal results for speaker posteriors. 

\subsection{Initialization of VB Iterations}
\label{init_heur}

Firstly, we adopt a heuristic approach in order to initialize segment posteriors similar to the study in \cite{kenny2010diarFA}. In this setup, instead of starting with two speakers, we randomly initialize the segment posteriors with three speakers. After running the VB algorithm, we compute the pairwise distances among the speakers using their corresponding mean vectors and take the most distant two speakers. Moreover, we iterate this procedure ten times and choose the final speaker pair among the most distant speakers of each iteration. Speaker pair which yields the furthest distant is chosen to be our starting point. We continue to the VB e-step iterations with these two speaker posteriors. As a distance metric we use cosine similarity and likelihood ratio scoring with the PLDA model \cite{brummer2010twocov,prince2007PLDA}. 


\subsection{Deterministic Annealing variant of Variational Bayes}
\label{init_DA}

DA is introduced to the VB method in order to avoid trapping in poor local optimal solutions. This process simply consists of introducing a temperature parameter, $\beta$, to the free energy for controlling the annealing process deterministically \cite{katahira2008DAVB}. The DA variant of update formulation in section \ref{vb_for_PLDA_ivec} can be adapted as follows: 
\begin{eqnarray}
\begin{split}
\log \tilde{q}_{ms}= &\beta(\mu_s^T\mathcal{L}\phi_m-\frac{1}{2}\mathrm{tr}(\mathcal{L}(C_s^{-1}+\mu_s\mu_s^T))) \\
                     &+\mbox{const}
\end{split}
\label{bete_q_ms}
\end{eqnarray}

\begin{eqnarray}
C_s=\beta(\Lambda+\sum_{m=1}^{M}q_{ms}\mathcal{L})
\label{beta_C_s}
\end{eqnarray}
\begin{eqnarray}
\mu_s=C_s^{-1}\beta(\Lambda\mu+\sum_{m=1}^{M}q_{ms}\mathcal{L}\phi_m)
\label{beta_mu_s}
\end{eqnarray}
By introducing the temperature parameter $\beta$ to the formulation, we attain a control on the convergence of the VB algorithm by lowering the precision $C_s$ of the speaker posterior distribution as seen in equation (\ref{beta_C_s}). 

\section{Experimental Setup and Results}
\label{sec:exp}

We use 20 dimensional static MFCC features. We use telephone part of the NIST 2004/2005/2006 SRE corpora in order to train gender-independent universal background model (UBM) of 1024 Gaussians. We train gender-independent i-vector model of rank 600 on the same dataset. We extract 600-dimensional i-vectors by using the sufficient statistics collected from the UBM in each segment.

\subsection{Segmentation}

After extraction of MFCC features, we use BIC based penalized likelihood ratio test to detect speaker change points. We check whether the data in the two sides of a candidate change point is better modeled with a single distribution or two. We use full covariance Gaussian distribution for modelling. This is the most widely used approach to speaker diarization for segmentation. Readers may refer to \cite{mit2004diarization} for detailed formulation and configurations.

\subsection{K-means clustering i-vector System}

This system is based on the work described in \cite{shum2011intraconv}. After extracting an i-vector for each speech segment in a given utterance, we apply principle component analysis (PCA) based projection. We choose the dimension of PCA-projected vectors for each utterance separately, so that 50\% of the energy is preserved. Then, we apply k-means ($K=2$) clustering to the projected i-vectors based on the cosine distance. 

\subsection{i-vector PLDA System}

In our proposed system, we apply linear discriminant analysis (LDA) to the segment i-vectors. After LDA, we apply whitening and unit length normalization before training the PLDA model. We use the same dataset with UBM training for training LDA and PLDA models. In speaker verification, a major source of intra-speaker variability is microphone and channel variations between utterances. For speaker diarization, we have a single session, and phonetic content variabilities are one of the major sources of variation between segment i-vectors of a given speaker. Hence, to obtain a better PLDA model for our task, we take a single utterance from every speaker in the training set. We use the i-vectors extracted from this full utterance, as well as from random cuts between 2 and 20 seconds extracted from it, in LDA and PLDA training. We observe a minor improvement compared to training on multi-session full utterances. 

\subsection{Viterbi re-segmentation}

After we complete the initial clustering step by using the VB algorithm, we conduct a frame-based Viterbi re-segmentation to improve the diarization result. We use the labels obtained from the initial clustering step to train 32 mixture GMMs for each speaker. We run the Viterbi algorithm, by fixed self-transition probability, over all speech frames with the two GMMs to obtain final alignments.


\subsection{Evaluation Protocol}
\label{eval_protocol}

The performance measurement of speaker diarization system is evaluated using diarization error rate (DER). This performance metric is calculated as alignment of reference diarization output with a system diarization output by summing up time weighted combination of: {\it Miss} - classifying speech as non-speech, {\it False Alarm} - classifying non-speech as speech and {\it Speaker Error} - confusing one speaker’s speech as from another \cite{mdeval}. The evaluation code ignores errors of less than 250ms in the locations of segment boundaries. We take the reference speech activity boundaries as given by using time marks from the speech recognition transcripts produced on each channel separately. Clearly, miss and false alarm errors are mainly caused by a mismatch of the reference speech activity detector and the diarization system output. For a more efficient metric in order to evaluate the effectiveness of our speaker diarization system based on the use of reference speech/non-speech boundaries, we set both miss and false alarm error rates to zero \cite{kenny2010diarFA, shum2011intraconv}.

\subsection{Results}

We use NIST SRE 2008 summed channel telephone data as test set. The dataset consists of 2215 conversations. Each conversation is approximately five minutes in duration ($\approx200$ hours in total) and involving just two speakers. In the experiments, we use 600-dimensional i-vectors to which we apply a dimensionality reduction procedure. For our system 150-dimensional LDA projection is employed and for the baseline system, we use utterance specific PCA projection keeping $50\%$ of the eigenvalue mass.

Table \ref{sys_comparison_base} shows the results of the baseline system (KM-PCA) as well as the results of our proposed system (VB-PLDA) which is initialized with two speakers and randomly generated segment posteriors. 

\begin{table}[h]\caption{Comparative results of baseline and proposed system. We randomly initialize $q_{ms}$ with two speakers for VB iterations.}\label{sys_comparison_base}
\centering
\begin{tabular}{|l|c|c|c|}
\hline
&KM-PCA&VB-PLDA\\ \hline
mean DER (\%) & 2.72 & 4.14  \\ \hline
$\sigma$ (\%)        & 5.83 & 9.16  \\ \hline
\end{tabular}
\end{table}
Table \ref{sys_comparison_init} shows the results obtained from our proposed system with two different heuristic initializations and a DA variant of VB. We use two metrics for initialization with cosine similarity (VB-COS) and PLDA log-likelihood ratio  (VB-LLR) described in section \ref{init_heur}. We apply four VB iterations in order to determine best two speaker models out of three for each ten attempts. For obtaining the results of DA variant of VB (DA-VB) system detailed in section \ref{init_DA}, we set initial value of temperature parameter as, $\beta_{init}=0.2$ and update as, $\beta_{new}=\beta_{current}\times1.05$ and continue to the VB iterations as long as $\beta_{new}<1$. By using DA, we obtain comparable performance to the cumbersome heuristic initialization methods.
\begin{table}[h]\caption{Comparative results of proposed systems with two different VB initializations and the DA variant of VB.}\label{sys_comparison_init}
\centering
\begin{tabular}{|l|c|c|c|c|}
\hline
&VB-COS&VB-LLR&DA-VB\\ \hline
mean DER (\%) & 2.18 & 2.19 & 2.28 \\ \hline
$\sigma$ (\%)        & 5.55 & 5.42 & 5.73 \\ \hline
\end{tabular}
\end{table}

\vspace{-0.1in}
\section{Conclusion}
\label{sec:conc}
Motivated by a previous study which utilizes factor analysis with a VB method \cite{kenny2010diarFA}, we develop a system that uses PLDA modelling with a VB method for inference in the speaker diarization problem. We successfully apply DA method to avoid the suboptimal heuristic initialization in VB. We obtain competitive performance as far as the study in \cite{shum2011intraconv} is concerned in our experiments.

Our future efforts will continue to apply proposed system to meeting and broadcast data involving an unknown number of speakers.

\section{RELATION TO PRIOR WORK}
\label{sec:prior}

In our proposed study we are inspired from a previous study \cite{kenny2010diarFA}, which exploits eigenvoice priors for VB. By our proposed system, we try to obtain a better modeling for the underlying distribution of the speaker factors of the i-vector in a probabilistic framework with the PLDA model which proved to be very successful in speaker recognition. In an another study \cite{prazak2011clusterPLDA}, PLDA is introduced in speaker diarization to compute the log-likelihood ratio as a substitute to BIC scores in the clustering stage. However, we use the PLDA model to represent segmental i-vectors and apply a VB approach for inference in this framework. Moreover, we introduce a formulation based on the DA variant of VB by which we overcome the initialization problem handled by a heuristic method in \cite{kenny2010diarFA}. 

\section{ACKNOWLEDGEMENTS}
\label{sec:acknow}
We would like to thank Patrick Kenny for giving the advice about using DA approach in VB. 

\bibliographystyle{IEEEbib}
\bibliography{references}

\end{document}